# Electrically tunable selective reflection of light from ultraviolet to visible and infrared by heliconical cholesterics

Jie Xiang[1], Yannian Li[1], Quan Li[1], Daniel A. Paterson[2], John M. D. Storey[2], Corrie T. Imrie[2],

Oleg D. Lavrentovich[1,*]

*1 Liquid Crystal Institute and Chemical Physics Interdisciplinary Program,
Kent State University, Kent, USA*
*2 Department of Chemistry, University of Aberdeen, AB24 3UE Scotland, UK*

**Abstract:** Cholesteric liquid crystals with helicoidal molecular architecture are known for their ability to selectively reflect light with the wavelength that is determined by the periodicity of molecular orientations. Here we demonstrate that by using a cholesteric with oblique helicoidal (heliconical) structure, as opposed to the classic "right-angle" helicoid, one can vary the wavelength of selectively reflected light in a broad spectral range, from ultraviolet to visible and infrared (360-1520 nm for the same chemical composition), by simply adjusting the electric field applied parallel to the helicoidal axis. The effect exists in a wide temperature range (including the room temperatures) and thus can enable many applications that require dynamically controlled transmission and reflection of electromagnetic waves, from energy-saving smart windows to tunable organic lasers, reflective color display, and transparent "see-through" displays.

Cholesteric liquid crystals with helicoidal molecular architecture are known for their ability to selectively reflect light with the wavelength that is determined by the periodicity of molecular orientations.[1] Resulting interference colours are highly saturated, they add like coloured lights and produce a colour gamut greater than that obtained with inks, dyes, and pigments.[2] The periodicity of the helical structure and thus the wavelength of the reflected light can be controlled by chemical composition and sometimes by temperature, but tuning with the electric field has been so far elusive. Here we demonstrate that by using a cholesteric with oblique helicoidal (heliconical) structure, as opposed to the classic "right-angle" helicoid, one can vary the wavelength of selectively reflected light in a broad spectral range, from ultraviolet to visible and infrared, by simply adjusting the electric field applied parallel to the helicoidal axis. The effect exists in a wide temperature range (including the room temperatures) and thus can enable many applications that require dynamically controlled transmission and reflection of electromagnetic waves, from energy-saving smart windows to tunable organic lasers, and transparent "see-through" displays. Since the material is non-absorbing and transparent everywhere except the electrically preselected reflection band, the effect can be used in creating multilayered structures with a dynamic additive mixture of colours.

Development of materials capable to dynamically control transmission and reflection of visible light and near infrared (IR) radiation is one of the most important directions of research with potential applications such as energy-saving smart windows,[3] transparent displays, communications, lasers,[4] etc. The most challenging problem is to formulate materials in which the transmission of light, or, more generally, electromagnetic radiation, can be performed dynamically and independently for different spectral bands, preferably by controlling reflection (as opposed to absorption). Among the oldest known materials capable of selective reflection of light are the so-called cholesteric liquid crystals (LCs) formed by chiral elongated organic molecules in a certain temperature range between a solid crystal and an isotropic melt.[2, 4-8] The molecules are packed into a periodic structure in which the local orientation of molecules, called the director $\hat{\mathbf{n}}$, rotates around a helicoidal axis, remaining everywhere perpendicular to it, **Figure 1a**. The pitch $P_0$ of the resulting right-angle helicoid is typically in the range (0.1-10) μm. The cholesteric separates the light traveling along the helicoidal axis into right-handed and left-handed circularly polarized components. One component, of the same handedness as the cholesteric, is reflected, and the other is transmitted. The selective reflection is observed in the spectral range $\Delta\lambda = (n_e - n_o)P_0$ determined by the pitch and the ordinary $n_o$ and extraordinary $n_e$ refractive indices. The band is centered at $\lambda_p = \bar{n}P_0$, where $\bar{n}$ is the average refractive index.

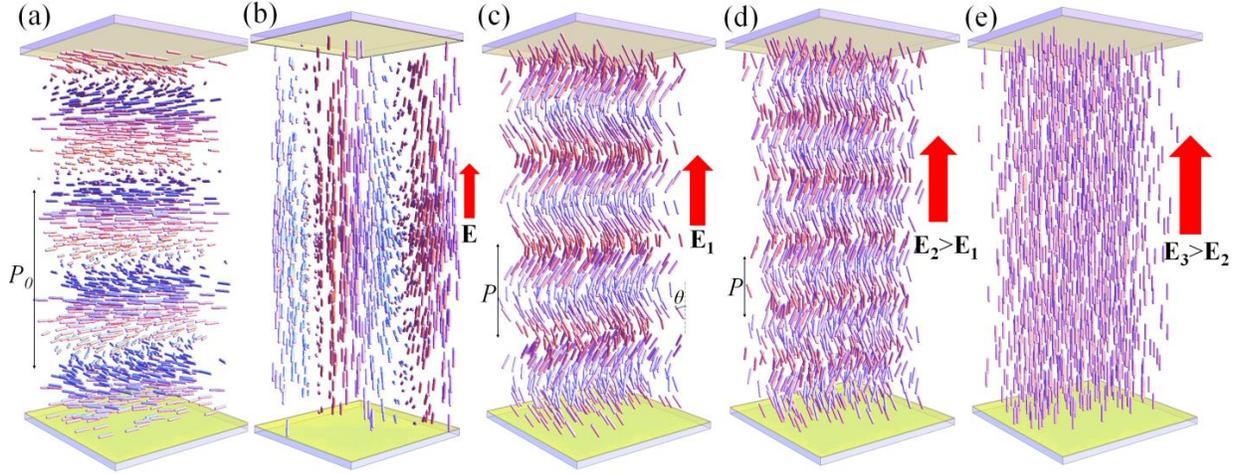

*Figure 1. Field induced behavior of cholesterics structures.* *(a) Right-angle helicoidal cholesteric with a large bend constant and a positive local dielectric anisotropy in a planar cell. (b) Sufficiently strong vertical electric field* **E** *realigns the cholesteric axis perpendicularly to itself, causing light-scattering fingerprint texture. (c) Heliconical structure in a cholesteric with a small bend constant and positive dielectric anisotropy stabilized by the vertical electric field* $\mathbf{E}_1$. *(d) The pitch* $P$ *and tilt angle* $\theta$ *of the field-induced heliconical state both decrease as the electric field increases,* $\mathbf{E}_2 > \mathbf{E}_1$. *(e) As the field increases further, to some* $\mathbf{E}_3 > \mathbf{E}_2$, *it unwinds the helical structure completely and forms a homeotropic nematic state. The figures are not to scale, as the experimental cell thickness is typically 20-50 times larger than the cholesteric pitch* $P_0$.

The pitch $P_0$ and thus the wavelength $\lambda_p$ of reflected light are sensitive to chemical composition and temperature, thus enabling applications such as temperature indicators and sensors of minute quantities of gases.[1, 9, 10] However, the most desirable mode to control light reflection, by an electric field applied parallel to the helicoidal axis, Figure 1a, has so far been elusive. The reason is that the field applied parallel to the axis, instead of changing the pitch while keeping the cholesteric axis intact to reflect light, rotates this axis perpendicularly to itself, as dictated by the dielectric anisotropy of the LC,[11] causing a light scattering structure called a "fingerprint texture", Figure 1b.

In this work, we demonstrate electrically controlled selective reflection of electromagnetic radiation in a broad spectral range, from ultraviolet (UV) to visible and IR, by using an oblique helicoid structure of cholesteric, Figure 1c, as opposed to the right-angle helicoid shown in Figure 1a. In the oblique helicoid, the director is tilted, making some angle $\theta < \pi/2$ with the helicoid axis. The electric field acting along the axis, realigns the molecules along itself and thus changes the pitch $P$ without reorienting the helicoid axis. The oblique helicoidal state was made stable in a broad temperature range including the room temperatures by designing the chemical composition of cholesteric mixtures.

The oblique helicoid state has been predicted by Meyer and de Gennes to occur in the applied electric field in a cholesteric in which the elastic constant $K_3$ of bend is much smaller than the elastic constant of twist.[12, 13] For a long time, the effect could not be explored because of lack of materials with the needed elastic properties. Recently, it was realized that the novel dimeric LCs, representing two rigid rod-like units connected by a flexible chain with an odd number of links, might have a small value of $K_3$, Ref.[14], which was confirmed experimentally.[15-17] The oblique helicoidal state of the corresponding cholesterics was observed in Raman-Nath diffraction experiments,[18] in which light propagates perpendicularly to the axis, and in Bragg reflection geometry,[19] but only at temperatures above 100 °C and short tuning range of the reflection band, not quite suitable for applications. In this work, we formulated a broad temperature range cholesterics with the small $K_3$ by mixing two dimeric LCs (1',7'-bis(4-cyanobiphenyl-4'-yl)heptane (CB7CB) and 1-(4-cyanobiphenyl-4'-yl)-6-(4-cyanobiphenyl-4'-yloxy)hexane (CB6OCB)), and a standard LC pentylcyanobiphenyle (5CB) (Merck). The mixtures were doped with a left handed chiral additive S811 (Merck) that determines $P_0$. Three mixtures were used, with composition CB7CB:CB6OCB:5CB:S811 (in weight units) being 30:20:46:4 (mixture M$_1$, cholesteric phase in the range (20-66.5) °C); 30.1:20:45.9:4 (M$_2$, 22-68 °C); and 29:20:49:2 (M$_3$, 21-69.5 °C). All mixtures demonstrated electrically tunable light reflection in the cholesteric phase at temperatures up to 45 °C; at higher temperatures, the effect disappears as $K_3$ in dimeric materials increases with temperature.[15-17] All data reported below were obtained at 25 °C.

In the experiments, the cholesteric is sandwiched between two glass plates with transparent indium tin oxide (ITO) electrodes. When a sufficiently strong electric field is applied, the material is switched into a uniform nematic with the director parallel to the field, **Figure 2a**. Such a state is dark when viewed between crossed linear polarizers. When the field is decreased, the LC shows a sequence of changing wavelength of reflection, from UV to visible blue, then green, orange, red, and, finally, near IR, Figure 2b-g. Below $0.7 \text{ V/}\mu\text{m}$, the LC transforms into the light scattering texture, Figure 2h.

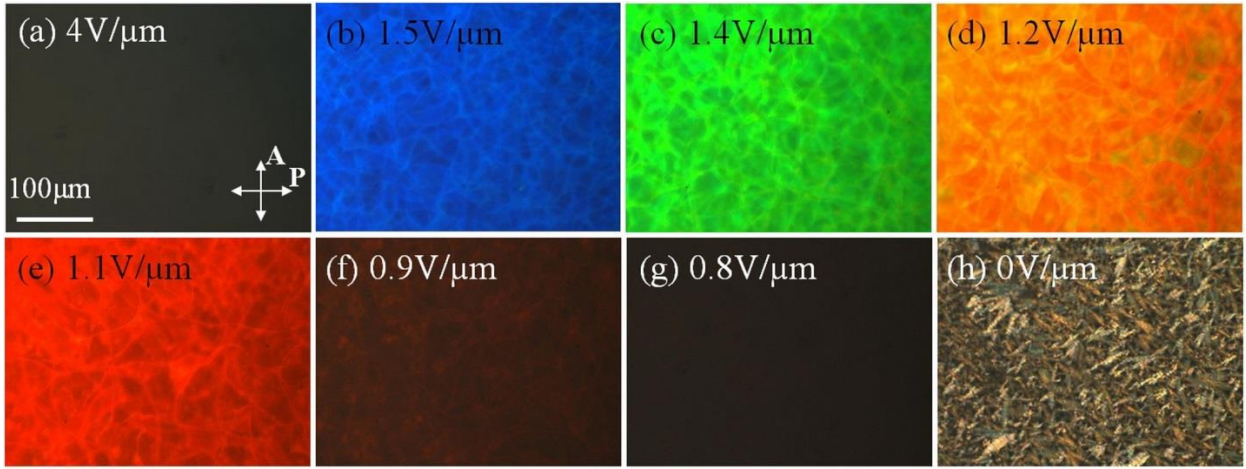

*Figure 2. Electric field induced textures in cholesteric mixture $M_1$. Polarizing optical microscope textures of field induced (a) unwound nematic; (b-g) heliconical states with reflected (b) blue, (c) green, (d) orange and (e) red colours, (f and g) two IR-reflective states; (h) fingerprint state. The RMS amplitude of the electric field is indicated on the figures.*

**Figure 3** presents reflection spectra, peak wavelength and bandwidth of reflection at various field strengths that further demonstrate a very broad range of controlled reflectance, from UV to IR, covering the entire range of visible light. As the field decreases, the peak wavelength shifts to IR, Figure 3a,b. In absolute units, the reflection coefficient $R$ was measured to be 41% for the reflection peak at 632 nm (field 1.1 $\text{V/}\mu\text{m}$). By controlling the chemical composition of the mixture, one can optimize the reflection efficiency for any given pitch, and adjust the bandwidth that narrows down at higher electric field, Figure 3b.

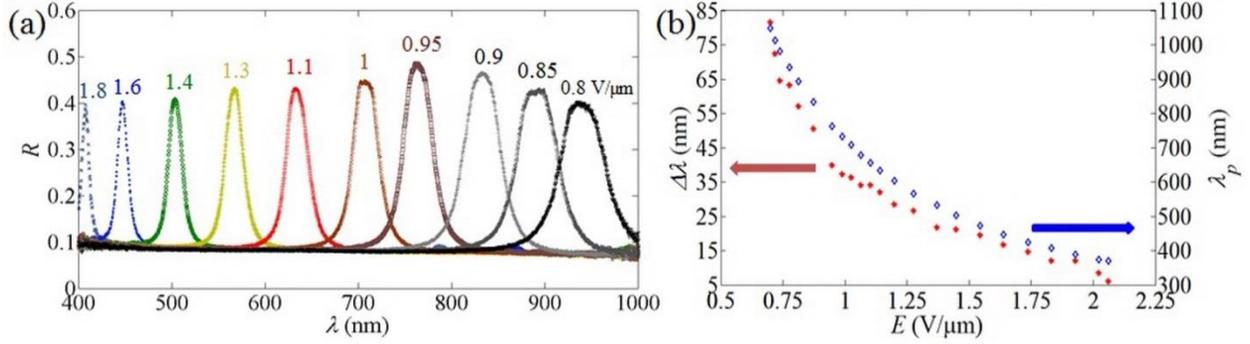

*Figure 3. Selective light reflection in $M_1$ cell.* *(a) Typical reflection spectra of $M_1$ cell for different amplitudes of the electric field, shown underneath the spectra in V/µm units. (b) Electric field dependencies of the wavelength and bandwidth of the selective reflection peak.*

An important goal in the field of smart windows is to block visible and near-IR light selectively and independently, by varying the applied voltage.[20, 21] Such a task can be performed by multilayered stacks of oblique helicoidal cholesterics with different concentration of chiral additive, since the materials are not absorbing. To demonstrate the principle, we stack the cells with short-pitch $M_2$ and long-pitch $M_3$ on top of each other, **Figure 4**. Depending on the applied fields, the stack produces a variety of states: a state (I) transparent in the visible region and reflecting in IR (a stronger voltage is applied to $M_2$ as compared to $M_3$ cell), Figure 4b; (II) transparent in IR and reflecting in the visible (stronger voltage applied to $M_3$ cell), Figure 4c; (III) reflecting in both visible and IR (moderate voltages), Figure 4d; (IV) transparent in both the visible and IR part (high voltages). Importantly, the electric field not only switches the two reflection bands on and off (as in standard cholesterics), it also tunes their spectral position, Figure 4a. Using the rich arsenal of techniques developed previously for standard cholesterics, one can broaden the reflected bands by polymerization of the material,[8, 11, 22, 23] or increase the reflectivity to 100% by using optical compensators.[2]

Experimental results on the field dependence of $\lambda_p$, Figure 3b, support the theoretical model[12] in which $P \propto 1/E$. For full description of the dependency $\lambda_p(E)$ one also needs to take into account a somewhat weaker field dependency of the refractive indices. For light propagating along the axis, the effective birefringence depends on the tilt angle $\theta$:
$\Delta n_{eff} = \dfrac{n_e n_o}{\sqrt{n_e^2 \cos^2\theta + n_o^2 \sin^2\theta}} - n_o$ . As demonstrated in Ref.[18], $\theta$ decreases as the field increases, so that $\Delta n_{eff}$ becomes smaller. This is why, for a given material, the reflectivity is typically weaker at high fields; it is also weaker at low fields, as the number of cholesteric layers in the cell of a fixed thickness decreases as the pitch increases, Figure 3a.

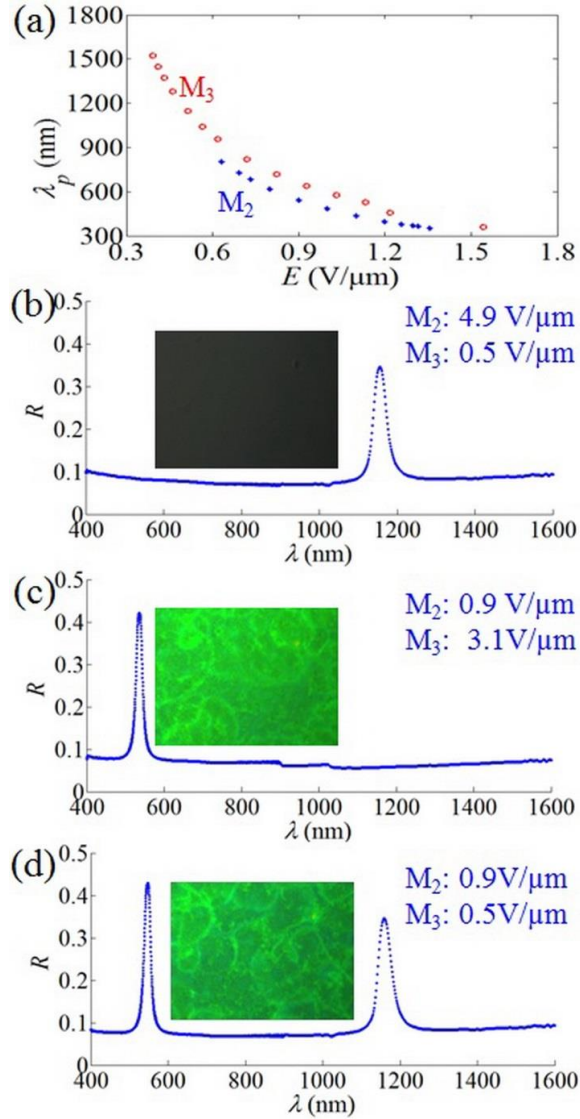

*Figure 4. Selective light reflection in double cell.* *(a) Reflection wavelength vs electric field for $M_2$ and $M_3$ cells. (b) State (I) transparent in visible and reflecting in IR. The electric field $4.9\,\text{V}/\mu\text{m}$ acts on $M_2$ cell and $0.5\,\text{V}/\mu\text{m}$ on $M_3$ cell; (c) state (II) transparent in IR and reflecting in visible; $0.9\,\text{V}/\mu\text{m}$ at $M_2$ cell and $3.1\,\text{V}/\mu\text{m}$ at $M_3$ cell; (d) State (III) reflecting in two different parts of spectrum; $0.9\,\text{V}/\mu\text{m}$ at $M_2$ cell and $0.5\,\text{V}/\mu\text{m}$ at $M_3$ cell. The textures are taken under a reflective optical microscope with crossed polarizers.*

When the light beam is not strictly perpendicular to the cell, the oblique helicoid produces reflection not only at $\lambda_p = \bar{n}P$, but also at $\Lambda_p = 2\bar{n}P$, since the molecules separated by $P/2$ are tilted in opposite directions.[24] In the presented experiments with normally incident beams of small divergence; the peak at $\Lambda_p = 2\bar{n}P$, although observable, was about 10 times weaker than the main peak at $\lambda_p = \bar{n}P$.

To change the colour, the pitch needs to adjust to the field, either through slippage at the bounding plates or through nucleation and propagation of dislocations; some of these defect lines are seen in Figure 2b-e. An important advantage of the heliconical cholesteric over similar structures in smectics is absence of positional order in molecular packing, which allows the system to adapt easily and reversibly to the changes of period; all the colour changes observed in our experiments are fully reversible and relatively fast. For example, the homeotropic state switches into the heliconical state with reflective red colour within 0.8 second, Figure S1.

To conclude, we describe an electrically tunable selective reflection of light in UV, visible and IR by the heliconical cholesteric state that exist in a broad temperature range including the room temperatures. The tunable structural colour is achieved in a simple sandwich geometry, in which the cholesteric is confined between two plates with transparent electrodes, thus implying a low cost and easy fabrication process. The colour change occurs over the entire electrode area that can be designed as a desired pattern suitable, for example, for "see-through" colour displays. The electrically tunable colours can be additionally controlled by the cholesteric composition and by employing reflective stacks. The effect can be tuned to practically any spectral regions in UV, visible and IR by chiral additives of different twisting power or concentrations. The regular right-angle cholesterics are known to provide structural colours of certain birds, beetles[25] and plants;[26] it would be of interest to explore whether the oblique helicoidal states occur in the natural world.

**Experimental**

The main component of the explored mixtures that yields the necessary smallness of $K_3$ is CB7CB.[14, 18] It shows a uniaxial N phase in the range (103-116)°C between the isotropic and the twist-bend nematic phase $N_{tb}$. At the field frequency 10 kHz, the dielectric permittivities parallel and perpendicular to the director were measured to be $\varepsilon_\parallel = 7.3$ and $\varepsilon_\perp = 5.9$, respectively; the dielectric anisotropy of the material is thus positive[14], so that the director prefers to align parallel to the electric field. At $106\,^{\circ}\text{C}$, the elastic constants are $K_1 = 5.7\,\text{pN}$, $K_2 = 2.6\,\text{pN}$ and $K_3 = 0.3\,\text{pN}$,[18] while the refractive indices are measured in our laboratory by the wedge cell technique[27] to be $n_e = 1.73 \pm 0.01$ and $n_e = 1.58 \pm 0.01$ (at $\lambda = 632\,\text{nm}$). CB6OCB is also of positive dielectric anisotropy; it shows a uniaxial N phase in the range (110-157)$^{\circ}\text{C}$.

The temperature was controlled by a hot stage LTS350 with a controller TMS94 (both Linkam Instruments) with $0.01\,^{\circ}\text{C}$ accuracy. All cells in the selective reflection experiments were addressed with the AC electric field of frequency 3 kHz (square wave). Flat cells were formed by glass plates with transparent ITO electrodes and alignment polyimide PI-1211 (Nissan); the thickness of cholesteric layers was $d = 50 \pm 2\,\mu\text{m}$. Selective light reflection was characterized by two complementary approaches. First, the field-induced colour changes were visualized under the polarizing microscope (Optiphot2-pol, Nikon) with two crossed linear polarizers, in the reflection mode, Figure 2. Second, we measured the reflection spectra of the cholesteric heliconical structure using Ocean Optics spectrometers USB2000 (visible) and NIRQuest256 (near-IR).

The absolute value of reflectivity coefficient $R$ at $\lambda_p = 632\,\text{nm}$ was determined by measuring the transmitted intensity $I_{ch}$ of a linearly polarized He-Ne laser beam ($\lambda = 632\,\text{nm}$) through the cell with the oblique helicoidal cholesteric and the intensity $I_i$ transmitted through the same cell with the isotropic phase of the material, $R = 1 - I_{ch}/I_i$. The obtained value $R = 0.41$ (or 41%) was used to normalize the values of $R$ measured by the spectrometers. The characteristic times of the electro-optic response was measured by recoding the transmittance change of circularly polarized light ($\lambda$=632nm), Figure S1.

**Acknowledgements**

This work was supported by NSF DMR- 1410378 and DMR-1121288. We thank V. Borshch for helping with preparation of illustrations, to Y. K. Kim for the help in experiments, V. A. Belyakov and S. V. Shiyanovskii for useful discussions.

* olavrent@kent.edu

**References**

[1]  P. G. de Gennes, J. Prost, *The Physics of Liquid Crystals*, Clarendon Press, Oxford **1993**.
[2]  D. M. Makow, C. L. Sanders, *Nature* **1978**, 276, 48.
[3]  B. Richter, D. Goldston, G. Crabtree, L. Glicksman, D. Goldstein, D. Greene, D. Kammen, M. Levine, M. Lubell, M. Savitz, D. Sperling, F. Schlachter, J. Scofield, J. Dawson, *Rev. Mod. Phys.* **2008**, 80, S1.
[4]  H. Coles, S. Morris, *Nature Photon.* **2010**, 4, 676.
[5]  N. Y. Ha, Y. Ohtsuka, S. M. Jeong, S. Nishimura, G. Suzaki, Y. Takanishi, K. Ishikawa, H. Takezoe, *Nature Mater.* **2008**, 7, 43.
[6]  W. J. Chung, J. W. Oh, K. Kwak, B. Y. Lee, J. Meyer, E. Wang, A. Hexemer, S. W. Lee, *Nature* **2011**, 478, 364.
[7]  M. Faryad, A. Lakhtakia, *Adv. Opt. Photonics* **2014**, 6, 225.
[8]  M. Mitov, *Adv. Mater.* **2012**, 24, 6260.
[9]  T. Ohzono, T. Yamamoto, J.-i. Fukuda, *Nature Commun.* **2014**, 5, 3735.
[10] Y. Nagata, K. Takagi, M. Suginome, *J. Am. Chem. Soc.* **2014**, 136, 9858.
[11] S.-T. Wu, D.-K. Yang, *Reflective liquid crystal displays*, Wiley, Chichester ; New York **2001**.
[12] R. B. Meyer, *Appl. Phys. Lett.* **1968**, 12, 281.
[13] P. G. de Gennes, *Solid State Commun.* **1968**, 6, 163.
[14] M. Cestari, S. Diez-Berart, D. A. Dunmur, A. Ferrarini, M. R. de la Fuente, D. J. B. Jackson, D. O. Lopez, G. R. Luckhurst, M. A. Perez-Jubindo, R. M. Richardson, J. Salud, B. A. Timimi, H. Zimmermann, *Phys. Rev. E.* **2011**, 84, 031704.
[15] R. Balachandran, V. P. Panov, J. K. Vij, A. Kocot, M. G. Tamba, A. Kohlmeier, G. H. Mehl, *Liq. Cryst.* **2013**, 40, 681.
[16] V. Borshch, Y. K. Kim, J. Xiang, M. Gao, A. Jakli, V. P. Panov, J. K. Vij, C. T. Imrie, M. G. Tamba, G. H. Mehl, O. D. Lavrentovich, *Nature Commun.* **2013**, 4, 2635.
[17] K. Adlem, M. Copic, G. R. Luckhurst, A. Mertelj, O. Parri, R. M. Richardson, B. D. Snow, B. A. Timimi, R. P. Tuffin, D. Wilkes, *Phys. Rev. E.* **2013**, 88, 022503.
[18] J. Xiang, S. V. Shiyanovskii, C. Imrie, O. D. Lavrentovich, *Phys. Rev. Lett.* **2014**, 112, 217801.
[19] J. Xiang, S. V. Shiyanovskii, Y. Li, C. T. Imrie, Q. Li, O. D. Lavrentovich, in *Liquid crystals XVIII*, Vol. 9182 (Ed: I.-C.Khoo), SPIE, San diego, USA 2014, 91820P.
[20] B. A. Korgel, *Nature* **2013**, 500, 278.
[21] A. Llordés, G. Garcia, J. Gazquez, D. J. Milliron, *Nature* **2013**, 500, 323.
[22] S. Y. Lu, L. C. Chien, *Appl. Phys. Lett.* **2007**, 91, 131119.
[23] V. T. Tondiglia, L. V. Natarajan, C. A. Bailey, M. M. Duning, R. L. Sutherland, D. Ke-Yang, A. Voevodin, T. J. White, T. J. Bunning, *J. Appl. Phys.* **2011**, 110, 053109.
[24] V. A. Belyakov, V. E. Dmitrienko, *Optics of chiral liquid crystals*, Harwood Academic Publishers, London, UK **1989**.
[25] V. Sharma, M. Crne, J. O. Park, M. Srinivasarao, *Science* **2009**, 325, 449.
[26] S. Vignolini, P. J. Rudall, A. V. Rowland, A. Reed, E. Moyroud, R. B. Faden, J. J. Baumberg, B. J. Glover, U. Steiner, *Proc. Nat. Acad. Sci. USA* **2012**, 109, 15712.
[27] J. Kedzierski, Z. Raszewski, M. A. Kojdecki, E. Kruszelnicki-Nowinowski, P. Perkowski, W. Piecek, E. Miszczyk, J. Zelinski, P. Morawiak, K. Ogrodnik, *Opto-Electron. Rev.* **2010**, 18, 214.

**Supplementary Information**

We determined the characteristic times of electrooptic response by recoding the transmittance change of the circularly polarized light through the cell and determining the levels of 10% and 90% of the maximum transmittance. The switching time is 0.8 second for switching from the homeotropic state (applied electric field $E = 5 \text{ V}/\mu\text{m}$) to the oblique helicoidal state with reflective red colour ($E = 1.1 \text{ V}/\mu\text{m}$), Figure S1.

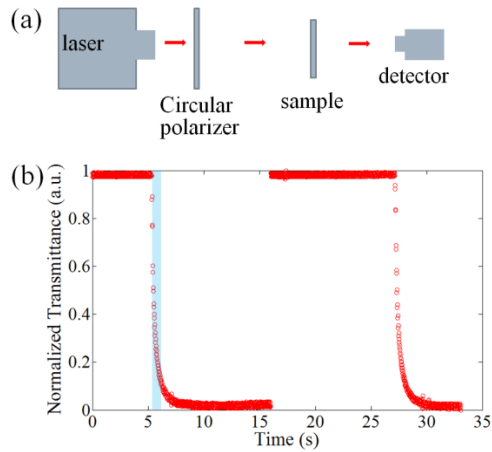

*Figure S1. Electro-optic response of cholesteric structures. (a) Schematic illustration of the experiment setup. (b) Dynamic process when the sample switched between hometropic state and heliconical state with red reflection colour ($\lambda_p$=632nm). The transmittance is 1 for homeotropic state, and 0 for heliconical state with red reflection colour ($\lambda_p$=632nm).*